**Polycrystal model of the mechanical behavior of a Mo-TiC$_{30\text{vol.\%}}$ metal-ceramic composite using a 3D microstructure map obtained by a dual beam FIB-SEM**

D. Cédat[a], O. Fandeur[b], C. Rey[a], D. Raabe[c]

[a] Departement Mechanical Soils, Structures and Materials, Ecole Centrale Paris, F-92295 Châtenay-Malabry, France
[b] CEA, DEN,DM2S, SEMT, LM2S, F-91191 Gif-sur-Yvette, France
[c] Department Microstructure Physics and Metal Forming, Max Planck Institut für Eisenforschung, Düsseldorf, Germany

**Abstract**

The mechanical behavior of a Mo-TiC$_{30 \text{ vol.\%}}$ ceramic-metal composite was investigated over a large temperature range (25°C to 700°C). High-energy X-ray tomography was used to reveal the percolation of the hard titanium carbide phase through the composite. Using a polycrystal approach for a two-phase material, finite element simulations were performed on a real 3D aggregate of the material. The 3D microstructure, used as starting configuration for the predictions, was obtained by serial-sectioning in a dual beam Focused Ion Beam (FIB)-Scanning Electron Microscope (SEM) coupled to an Electron Back Scattering Diffraction system (3D EBSD, EBSD tomography). The 3D aggregate consists of a molybdenum matrix and a percolating TiC skeleton.

As most BCC metals, the molybdenum matrix phase is characterized by a change in the plasticity mechanisms with temperature. We used a polycrystal model for the BCC material, which was extended to two phases (TiC and Mo). The model parameters of the matrix were determined from experiments on pure molybdenum. For all temperatures investigated, the TiC particles were considered as brittle. Gradual damage of the TiC particles was treated, based on an accumulative failure law that is approximated by an evolution of the apparent particle elastic stiffness. The model enabled us to determine the evolution of the local mechanical fields with deformation and temperature. We showed that a 3D aggregate representing the actual microstructure of the composite is required to understand the local and global mechanical properties of the studied composite.

## 1. Introduction

The aim of this study is to understand the mechanical behaviour and the damage evolutions of a metal-ceramic composite (Mo-TiC$_{30\text{vol.\%}}$) as a function of deformation and temperatures ranging from 800°C up to 1000°C.

For compression tests performed at high temperature, damage observations cannot be monitored and simulations are needed at grain scale, to get a better understanding of damage initiation..

In a preceding study, Cédat et al. [4] simulated the local mechanical fields using a 3D aggregate that was obtained by a random stack of layers of columnar grains. These grains were characterized in terms of crystal orientations that were determined by EBSD. When using this aggregate, the polycrystal model failed to describe the mechanical behavior of the composite. Different effects might be responsible for the discrepancy between experiment and predictions, namely, microstructural size effects, higher complexity of the required constitutive laws of the involved phases, and the spatial distribution (connectivity, and topology) of the TiC brittle phase.

The current paper addresses this problem of properly predicting the micromechanical behavior of complex metal-ceramic composites in the 5 subsequent sections.

In section 2, investigations (using tomography experiments) reveal a percolation of the TiC phase. To take into account such phase percolation within the computation of the local mechanical fields, a new representative aggregate has been built, using tomographic microstructure data obtained by a dual beam FIB-SEM experiment.

The used polycrystal model, implemented in a Finite Element code, is described in section 3. Damage is represented by a decrease of TiC effective elastic modulus, implemented in the model.

Section 4 presents the meshing and the boundary conditions applied to the 3D EBSD aggregate. New model parameters for Mo are determined by an inverse method. Section 5 deals with the results of the mechanical predictions for the composite. Section 6 draws conclusions.

## 2. Material and experimental procedures

The investigated material was obtained by powder metallurgy synthesis [2], [3] (Hot Isostatic Pressing (HIP) at 1600°C). No homogenization heat treatment was performed after HIP processing.

The composite microstructure was characterized in a previous paper [4] using different chemical and physical methods. According to Cédat [1], the composite exhibited three phases: Molybdenum (Mo), Titanium Carbide (TiC) and a (Mo,Ti)C phase which is a face-centered cubic (FCC) structure with lattice parameters close to those of TiC (Fig.1). The hard particles (TiC) showed a core/shell (or core/rim) structure, with molybdenum as binder phase. The third phase, identified as TiC-Mo$_{15at.\%}$ had the same structure as TiC. In this paper, the mechanical behavior of the third phase is assumed to be equivalent to that of TiC.

Molybdenum

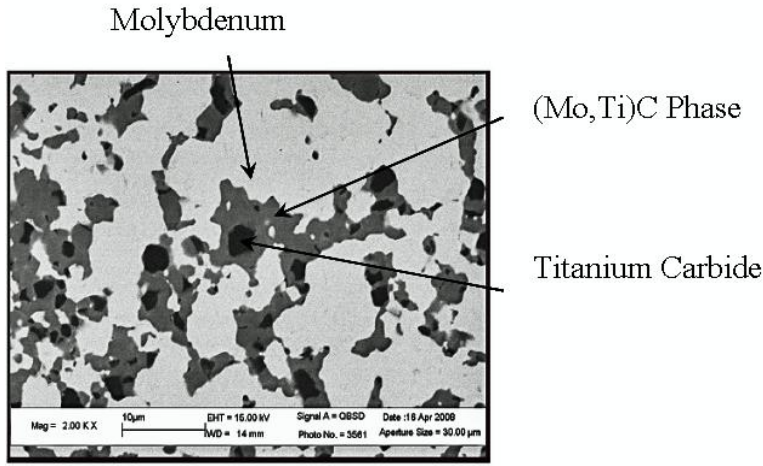

(Mo,Ti)C Phase

Titanium Carbide

*Fig.1. SEM picture showing the microstructure of the Mo-TiC$_{30\%vol}$ composite revealing three types of phases, namely, Molybdenum (in light grey), TiC (in black), and (Mo,Ti)C (in grey) [1].*

### 2.1 Chemical composition

The chemical composition of the powders is given in tables 1 and 2.

| C | O | N | Ca | Co | W | Ni | Al | Fe | S |
|---|---|---|----|----|---|----|----|----|---|
| 19.23 | 0.6126 | 0.0279 | 0.002 | 0.032 | 0.39 | <4E-4 | 0.0014 | 0.0061 | 0.0019 |

Table 1. TiC composition (%mass).

| Mo | O | Fe | K |
|----|---|----|---|
| 99.98 | 0.0620 | 9ppm | 29ppm |

Table 2. Molybdenum composition (% mass)

### 2.2 X-ray tomography results

Using high-energy X-ray tomography, a 3D investigation of the material was performed at the European Synchrotron Radiation Facility (ESRF, Grenoble, France). This characterization technique proceeds via taking a sequence of recordings of a sample at different angular positions. Then, an appropriate tomography reconstruction algorithm builds a 3D map of the microstructure based on the inversion of the X-ray attenuation coefficient projection paths that penetrated the material upon synchrotron illumination. Fig.2 shows a 3D image of the microstructure of the Mo-TiC material obtained with the highest spatial resolution (0.28 μm) available on the ID19 beam line (X-ray energy: 65 keV) of the ESRF equipment.

Fig.2 represents the set of all points in the carbide phase (purple) that can be linked together to the bottom and top faces of the parallelepiped specimen (30 X 30 X 30μm³), by paths entirely contained in the carbide phase. The microstructure of the composite revealed a percolating skeleton of carbides, embedded in a molybdenum metallic matrix, with a high 3D connectivity of the carbide phase.

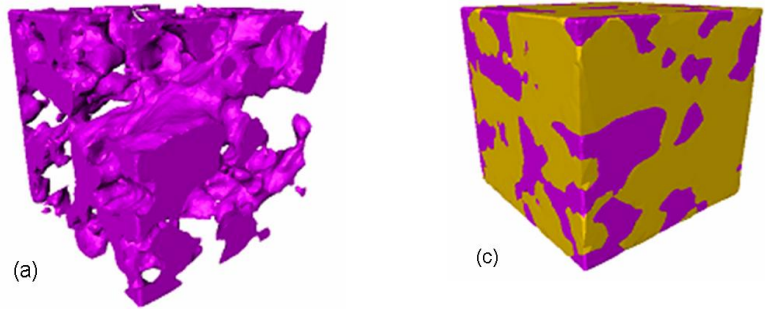

*Fig.2: 3D representation of the phases obtained by X-ray tomography. (a) TiC , (b) composite. Specimen size: (30 X 30 X 30 µm³)*

## 2.3   3D aggregate mapping via 3D EBSD

To accurately compute the local mechanical fields, a new representative 3D aggregate has to be designed, taking into account the observed percolation of the TiC phase. Our approach to 3D EBSD was inspired by the works of Uchic et al. [5], Zaefferer et al. [6], Bastos et al. [7] and Konrad et al. [8], as well as by the 3D texture measurements, via synchrotron radiation published by Larson et al. [9] and Yang et al. [10]. The 3D EBSD experiments were conducted at the Max-Planck Institute (Düsseldorf), using a joint high-resolution field emission SEM/EBSD set-up together with a FIB system in the form of a Zeiss cross-beam 3D crystal orientation microscope. Details are given in [9] [10].

The analysis method applied in this study, involves highly precise and fully automated serial sectioning via a Ga$^+$FIB and the subsequent mapping of the crystallographic texture in each of those layers using high-resolution EBSD. For more details on this method, see Zaafarani et al [11] and Zaefferer et al [12].

The spatial resolution of the 3D orientation and phase pixel information retrieved amounts to 50 x 50 x 50 nm³. The ion beam did not create noticeable damage, i.e. no serious deterioration of the EBSD pattern quality could be observed after milling. This stability of the material, when exposed to an ion beam, is attributed to its high melting point and the absence of phase transformations. The EBSD measurements were carried out sequentially in each layer after FIB serial sectioning at a 300nm step size between the abutting layers. The entire process of alternating FIB sectioning and EBSD measurements was carried out for a set of 22 subsequent layers for the reconstruction of the whole aggregate. Fig.3 shows SEM images of two different serial sections and the entire 3D microstructure reconstructed as a 3D map.

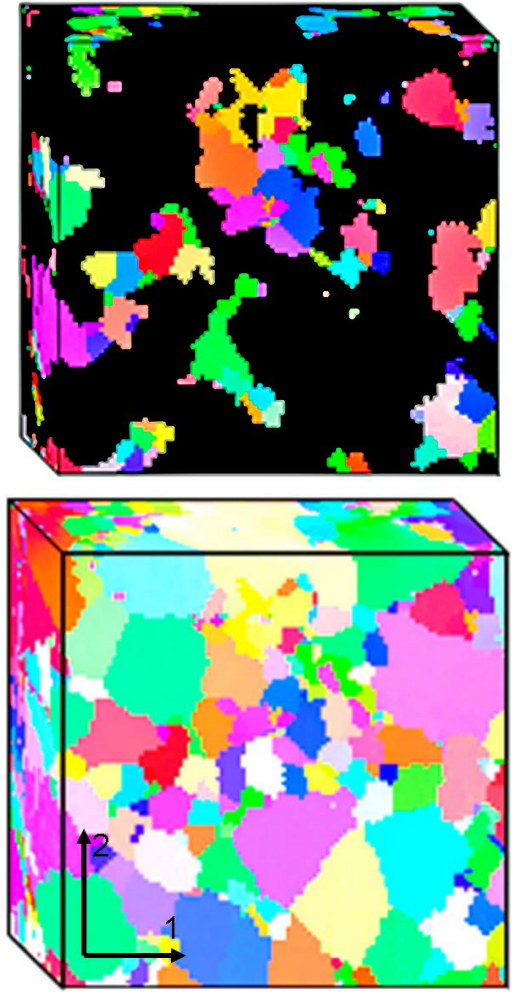

*Fig.3: 30 X 30 X 6.6 μm³ aggregate obtained by 3D electron orientation microscopy (3D EBSD). (a) TiC, (b) aggregate. The colour code is given by the crystallographic standard triangle. For simulations, the loading direction is parallel to $\bar{2}$ axis.*

## 2.3 Experimental analysis of the mechanical behavior of the composite and of the Molybdenum phase

*Composite*: The composite compression curves are given in Fig.4 for temperatures between 25°C and 700°C. The specimen are cylinders with a diameter D=12mm and cylinder length L=18mm. The initial strain rate upon loading was about $\dot{\varepsilon} = 5.10^{-4} \ s^{-1}$.

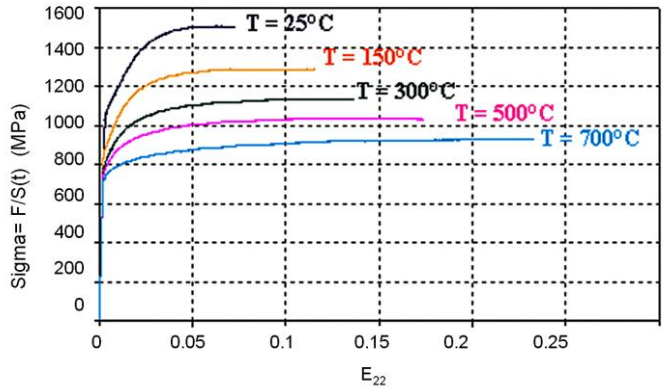

*Fig.4: Composite behavior compression experimental tests; sigma: tensile stress; F: force; S(t): cross section at each time interval ; $E_{22}$ : strain*

The compression curves show a linear elastic stage, followed by a linear hardening stage, and finally a weak strain hardening regime. The two strain hardening regimes change as a function of the temperature and deformation. The extension of the linear stage decreases with increasing total deformation. The strain hardening decays with increasing temperature.

Fracture surface observations reveal that at 25°C, the two phases are characterized by a brittle behavior and that at 300°C, the molybdenum matrix is ductile while the TiC phase fails in a brittle manner. The mechanical properties of pure molybdenum had been previously studied by Cédat et al [1] using two kinds of tensile tests, namely, strain rate change tests ($5.10^{-4}$ and $5.10^{-2}$ s$^{-1}$) over a temperature ranging from 25 to 700°C and temperature change tests. A transition in the polycrystal mechanical mechanisms was observed above 150°C. For T>300°C, the viscoplastic flow weakly varied with temperature. These results, confirmed by TEM observations, suggested that molybdenum (like most BCC metals) reveal different plasticity properties at low and high temperatures. In BCC transition metals, screw dislocations are less mobile than edge dislocations. Screw dislocation motion proceeds via thermally activated kink pair formation and expansion. Optical microscopy (OM) and SEM observations [13] performed on molybdenum surfaces as well as on fracture surfaces after tensile testing, confirmed these results. For T= 25°C, molybdenum rupture occurs by grain boundary decohesion. The studied fracture surfaces reveal features of brittle failure. For T=200°C, grain boundary decohesion and plastic glide is observed. For T=300°C, deformation occurs by plastic glide and the fracture surface is characterized by ductile rupture features.

According to these observations, we concluded in our earlier work that the plasticity mechanisms of this material change in the temperature range around 200°C. One should note that this transition temperature $T_0$ is different from the brittle-to-ductile transition temperature of molybdenum (which is close to 25°C).

## 3. Polycrystal model

In order to determine the local mechanical fields in the different phases and especially the location and origin of damage in the TiC phase, a previous model [14] was upgraded and numerical simulations were conducted. The used crystal plasticity model was developed in the framework of large transformations [15, 16, 17, 178, 19, 20] (small elastic distortion but large lattice rotation) and was implemented in the Abaqus finite element code, using an UMAT User Subroutine.

The polycrystal is composed of different grains, each one having its own size and crystallographic orientation. Each grain is considered as a single crystal that is characterized by a constitutive law that is presented below.

This polycrystal model was developed for BCC metals or alloys, taking into account temperature effects as will be outlined in more detail below. The two plasticity mechanisms controlling plastic strain hardening of molybdenum are double kink formation (low temperature) and interactions between mobile and forest dislocations (high temperature). In this paper, we summarize the polycrystal model extended to multi-phase composites. For more details, see Libert [18], Libert et al. [14].

### 3.1 Single crystal constitutive law for the molybdenum matrix

For the BCC crystal structure of the molybdenum matrix, plastic glide occurs on the 24 slip systems {110} <111> and {112} <111>. The chosen viscoplastic law can describe the competition between lattice friction and forest hardening. According to Kubin et al. [21], Rauch [22] and Tabourot et al. [23], the slip rate on a given slip system (hereafter referred to by the index (s)), is given by:

$$\dot{\gamma}^s = \dot{\gamma}_0 \, exp \left[ - \frac{\Delta G(\tau_{eff}^s)}{kT} \right] sgn(\tau^s) \qquad (1)$$

Where $\Delta G(\tau_{eff}^s)$ is the energy of activation for crystallographic slip of lattice dislocations, which is a function of the effective shear stress $\tau_{eff}^s$. $\tau^s$ and $\dot{\gamma}_0$ are the resolved shear stress on system (s) and the reference shear rate, respectively.

When T<$T_0$, the shear rate $\dot{\gamma}_0$ depends on the geometry of the segments of the screw dislocations. The energy of activation is then associated with the double kink mechanism and it is given by a phenomenological description, as proposed by Kocks et al. [24]:

$$\Delta G \left( \tau_{eff}^s \right) = \Delta G_0 \left( 1 - \left( \frac{\left| \tau_{eff}^s \right|}{\tau_R} \right)^p \right)^q \qquad (2)$$

Where p and q are parameters describing the energy associated with the Peierls barrier. In Eq.2, $\tau_R$ is the effective stress that is required to create a double kink at T=0 K, when the contribution of the thermal activation is equal to zero. This description assumes that the glide of dislocations on a system (s) is activated when $\tau_{eff}^s$ is larger than $\tau_R$.

## 3.2 Single crystal strain hardening law for the molybdenum matrix

The strain hardening evolution as a function of temperature depends on the interactions between mobile dislocations with lattice friction and/or forest dislocations. Lattice friction generates short range stresses described by the effective stress $\tau_{eff}^s$. The contributions described by $\tau_{int}^s$ and $\tau_0$ are shear stresses and stand for the interaction between mobile and latent dislocations. Thus, the yield stress required to activate the glide system (s) is equal to the sum of the three contributions:

$$\tau^s = \tau_0 + \tau_{eff}^s + \tau_{int}^s \qquad (3)$$

By considering a segment of a screw dislocation pinned by two obstacles, submitted to the lattice friction ($\tau_0$), Rauch [22] proposed the following expression of the internal stress:

$$\tau_{int}^s = \frac{(\mu b)^2 \sum\limits_{u=1,24} a^{su} \rho^u}{\tau^s - \tau_0} \qquad (4)$$

$a^{su}$ is a component of the dislocation interaction matrix as proposed by Franciosi [25], that is due to the interaction strength between the slip systems (s) and (u) (self and latent hardening). Latent hardening experiments [25], performed on single crystals under tensile testings, revealed that the $a^{su}$ parameters depend on the strain values: they increase up to 0.5% -1% straining, then reach asymptotic values. In this paper, we use the asymptotic values. $\rho^u$ is the dislocation density on the glide system (u). When merging expressions (3) and (4) and solving the resulting equation, one obtains:

$$\tau^s = \tau_0 + \frac{\tau_{eff}^s}{2} + \frac{1}{2} \sqrt{\left[ \left( \tau_{eff}^s \right)^2 + 4(\mu b)^2 \sum\limits_u a^{su} \rho^u \right]} \qquad (5)$$

The general expression of the hardening law (5) is appropriate to describe a continuous evolution of the shear stress as a function of the temperature:

In the case of low temperature behavior (T<$T_0$), plasticity is governed by the reduced mobility of the screw dislocations and the double kink mechanism hence, $\tau_{int}$ is negligible compared to $\tau_{eff}$. Eq.5 leads to:

$$\tau^s = \tau_{eff}^s + \frac{(\mu b)^2 \sum\limits_{u=1,24} a^{su} \rho^u}{\tau_{eff}^u} \qquad (6)$$

In the case of high temperature, (T>$T_0$), plasticity mainly depends on forest hardening and $\tau_{eff}$ is negligible compared the two other terms. Eq.5 leads to:

$$\tau^s = \tau_0^s + \mu b \sqrt{\sum\limits_{u=1,24} a^{su} \rho^u} \qquad (7)$$

At low temperature, plastic deformation occurs according to two non equivalent slip system families <111>{110} and <111>{112}.

### 3.3 Dislocation density evolution law

The dislocation density evolution law is a generalization of the relation proposed by Estrin and Mecking [26]. For each slip system, Eq. 8 describes the evolution of the 24 dislocation densities with the strain:

$$\dot{\rho}^s = \frac{|\dot{\gamma}^s|}{b} \left[ \frac{1}{D_{grain}} + \frac{\sqrt{\sum_{u \neq s} \rho^u}}{K(T)} - g_c(T)\rho^s \right] \tag{8}$$

This expression is derived from the balance between dislocation accumulation (Orowan's relationship) and dislocation annihilation which induces softening. $K(T)/\sqrt{\sum_{u \neq s} \rho^u}$ is the dislocations mean free path which increases with decreasing temperature. As the evolution of the dislocation density is weak for low temperature, we assume that $K(T)$ increases with decreasing temperature. The annihilation of dislocations is controlled by the mean dislocation-dislocation spacing $g_c$, the temperature dependence of which is expressed here in terms of an Arrhenius law:

$$g_c(T) = g_{c0} \exp\left[ -\frac{E_{gc}}{k_B(T)} \right] \tag{9}$$

According to Eq.9, two dislocations of the same system (s) may undergo annihilation, as soon as they are separated by a spacing below $g_c$. Similar dislocation-based constitutive laws for crystal plasticity finite element models were applied, as suggested by Ma et al. [17, 27].

### 3.4 Damage law for titanium carbide particles

Damage, in the compound, starts by initiation of cracks within the brittle carbide phase. A simple criterion that rests on the local stress in the brittle phase is introduced in the polycrystal model. The failure strength of ceramics under compression being approximately 10 to 15 times larger than the tensile strength [28], the critical stress leading to damage initiation is given by the following approximation:

$$\sigma_c = \sigma_{tensile} \approx \frac{\sigma_{compressive}}{15} \text{ with } \sigma_{tensile} = \frac{K_{IC}}{\sqrt{\pi a}} \tag{10}$$

$K_{IC}$ is the toughness factor and $a$ the mean size of the crack.

The results of the simulations show that the carbide phase is mainly under tension when the composite is mechanically loaded, while the molybdenum phase is under compression. For the carbide phase, the chosen damage criterion is formulated as a measure of the gradual accumulation of damage. It is based on the evolution of the elastic law:

$$\tilde{\tilde{C}}(t) = \tilde{\tilde{C}}_0[1 - D(t)] \tag{11}$$

with
$$D(t) = \frac{\beta}{\alpha}[1 - \exp(-\alpha(t - t_0))] \tag{12}$$

where $t - t_0 = \varepsilon / \dot{\varepsilon}$, and $\varepsilon$ and $\dot{\varepsilon}$ are the local strain and local strain rate at a given point, respectively.

$\tilde{\tilde{c}}_0$ is the initial elastic modulus tensor and $\tilde{\tilde{c}}_{(t)}$ is "the effective modulus" that gradually decays as a function of the strain rate and of the strain. $\alpha$ and $\beta$ are parameters that are arbitrarily chosen in order to create an abrupt decrease of the Young modulus due to micro-cracks formation. A positive critical stress value $\sigma_C$ is determined from Eq.10. At each increment of the simulation, the criterion (Eq.10) is tested for all elements of the TiC meshing. As soon as the criterion $\sigma \geq \sigma_c$ ($\sigma$ is the local equivalent stress) is verified within an element, its effective modulus is decreased as given by Eq.11, 12. For D(t) =1 (C(t)=0), the element is equivalent to an "hole". Changes of some elements stiffness of the TiC phase, contribute to that of the overall compound.

## 4. Boundary conditions, meshing and model parameters

The polycrystal model was implemented in the Abaqus$^{TM}$ finite element code using an UMAT subroutine. Most of the molybdenum parameters are determined on the basis of the experimental curves, by using an inverse fitting method using a coupling ([18, 20]) between the Sidolo$^{TM}$ and Abaqus$^{TM}$ software packages. The initial density of the dislocations before straining was determined by TEM. The parameters used for the molybdenum are given in tables 3 and 4.

*Aggregate*. The polycrystalline aggregate is shown in Fig.3. The Finite Element meshing is derived from the grid corresponding to the 3D EBSD (0.3 μm step) measurements. It is built by 220,000 elements (C3D8). Each element corresponds to an original material square pixel that is 0.3 μm wide.

The crystallographic microstructure properties of the two phases are described in terms of the Euler angles for each pixel of the two phases in the mesh, the TiC elastic properties, and the molybdenum constitutive laws parameters. The connectivity among the two phase sets and the percolation of the TiC are thus automatically taken into account.

*Boundary conditions*: The considered aggregate corresponds to 1/8 of the whole computed structure. For symmetry reasons, the lower face of the crystalline aggregate, shown in Fig. 3, is loaded in compression mode normal to direction $\vec{2}$. The displacement is $u_2$=0 on the lower face. The other boundary conditions are $u_3$=0 on the face normal to the $\vec{3}$ direction, the parallel surface is free. The $u_2$ applied displacement corresponds to a strain rate of about $5.10^{-4} s^{-1}$.

| $\tau_0$ | $a^{uu}=a^{su}$ | $\rho_0$ | $g_{c0}$ | $E_{gc}$ | | $D_{grain}$ |
|---|---|---|---|---|---|---|
| 125 MPa | 0.01 | $10^{-11} m^{-2}$ | 14nm | $2.17 \times 10^{-2}$ eV | | 3 μm |
| $\tau_R$ | $\dot{\gamma}_0$ | $\Delta G_0$ | p | q | | |

| | | | | |
|---|---|---|---|---|
| 498 MPa | $10^{-1}$ | 1.1 eV | 0.2 | 1.5 |

Table 3. *Set of parameters identified for molybdenum by Cédat et al [1]*

In addition to this set of parameters, this new investigation has identified the parameters depending on temperature by inverse method on the base of experimental tensile curves performed at 25°C to 700°C.

| Molybdenum parameters depending on temperature | | | | |
|---|---|---|---|---|
| T | $\Delta G(T)$ (eV) | $\tau_0$ (MPa) | K | $g_c$ (nm) |
| 25°C | 0.35 | 125 | 440 | 20 |
| 150°C | 0.5 | 70 | 80 | 12 |
| 300°C | 0.97 | 70 | 50 | 26 |
| 500°C | 1.18 | 70 | 50 | 30 |
| 700°C | - | 70 | 30 | 36 |

Table 4. *Set of temperature dependent parameters identified for molybdenum*

$\tau_0$ : shear stress due to lattice friction (Eq.3)

$a^{su}$: component of the interaction matrix between dislocation (s) and (u)  (Eq.5,6,7)

$\rho_0$ : initial dislocation density

$g_{c0}$: material parameter of the dislocation-dislocation annihilation law (Eq. 9)

$Eg_c$: material parameter of the dislocation -dislocation annihilation law (Eq.9)

$D_{grain}$: grain size

$\tau_R$ : effective stress required to create de double kink (Eq.2)

$\dot{\gamma}_0$ : shear rate reference (Eq.1)

$\Delta G_0$ : material parameter associated to the energy of activation of slip system (Eq.2)

p and q :parameters describing the energy associated with the Peierls barrier (Eq.2)

$\Delta G(T) = \Delta G(\tau_{eff})$: slip systems energy of activation (Eq.1)

K(T): material parameter associated to the dislocation mean free path (Eq.8)

$g_c$: dislocation-dislocation annihilation distance  (Eq.8,9)

The titanium carbide elasticity is assumed isotropic. Its features are given in table 5.

| E (GPa) | $\nu$ | $\alpha$ | $\beta$ | $\sigma_c$ (MPa) |
|---|---|---|---|---|
| 440 | 0.19 | 0.013 | 0.013 | 250 |

Table 5: Parameters of titanium carbide (TiC) for the proposed model

E is the Young modulus, $\nu$ the Poisson coefficient, $\sigma_c$ the rupture critical parameter in tension.

According to our observations of fracture surfaces, we assume a perfect cohesion between TiC and Mo matrix.

## 5. Results and discussion

For a pure molybdenum metal representing the polycrystal matrix, the simulation results delivered by the current polycrystal model were already published in [4]. The compression test simulations were in good agreement with the experiments results.

## 5.1 Mo-TiC composite results for T=25°C

The numerical and experimental curves are compared in Fig.5.

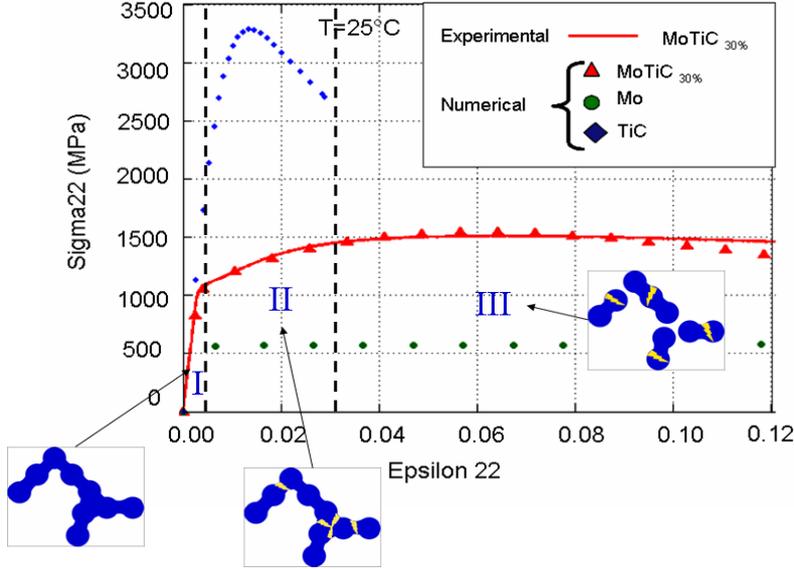

*Fig.4: Composite behavior compression experimental tests; sigma: tensile stress; F: force; S(t): cross section at each time interval ; $\varepsilon_{22}$ : strain.*

*Fig.5: Mo, TiC and Mo-TiC$_{30\%}$ numerical and experimental compression curves (T=25°C); sigma22: compression stress; epsilon22: strain. Scheme of TiC damage.*

At room temperature, a very good agreement is observed between the numerical (Fig. 5) and experimental curves (Fig. 4), up to the maximum stress. Compared to our first simulations [1] that were carried out on crystalline aggregates (constructed as superposition of random layers of columnar grains), a large improvement is observed in the current full field 3D predictions. This improvement is attributed to the more realistic 3D mapping of the incipient microstructure where both, percolation and damage of the TiC phase are taken into account.

For each phase, the computed local stress averages $\langle \sigma_{22} \rangle$ versus the applied strain $E_{22}$ are represented by dotted lines. The molydenum curve presents a perfect plastic behaviour, whereas TiC shows a non linear behaviour for $0.5\% \leq E_{22} \leq 1.3\%$, following by a maximum, then a decrease. The perfect plastic behaviour of the matrix corresponds to $\langle \sigma_{22} \rangle = 590\ MPa$ (close to the Mo pure metal behaviour). The matrix yield stress is about 550MPa and the constant stress corresponding to the plateau is equal to $\Sigma_{22}$ =620MPa up to E22=0.15. The non linear TiC curve corresponds to a decrease of the effective elastic modulus induced by damage.

The computed maps (Fig.6) represent the evolution of distribution of local strain $\varepsilon_{22}$ in the grains of Mo and TiC and of damage within the TiC particles. The local strains ($\varepsilon_{22}$) distribution is homogeneous in TiC grains and heterogeneous in large molybdenum grains. In TiC the damage parameter D(t) (introduced in eq. 10-12), increases with applied deformation and locally reach 1 for an applied strain $E_{22} \cong 10\%$. The value D=1 corresponds to rupture of the particles.

According to our numerical results, we can propose for damage the following scheme: during stage I (elastic stage), the TiC skeleton supports most of the deformation. At the end of stage I and up to 3% compression strain rate (stage II), some bridges of TiC (normal to the compression axes) are submitted to local tensile stress. As soon as the critical stress criterion (eq.10) is verified, some bridges between TiC particles are damaged. During stage II, the damage parameters D increases and reaches 1 (rupture of the bridge) for $E_{22} \cong 10\%$. In stage III, the composite behaves as a matrix with large disconnected particles.

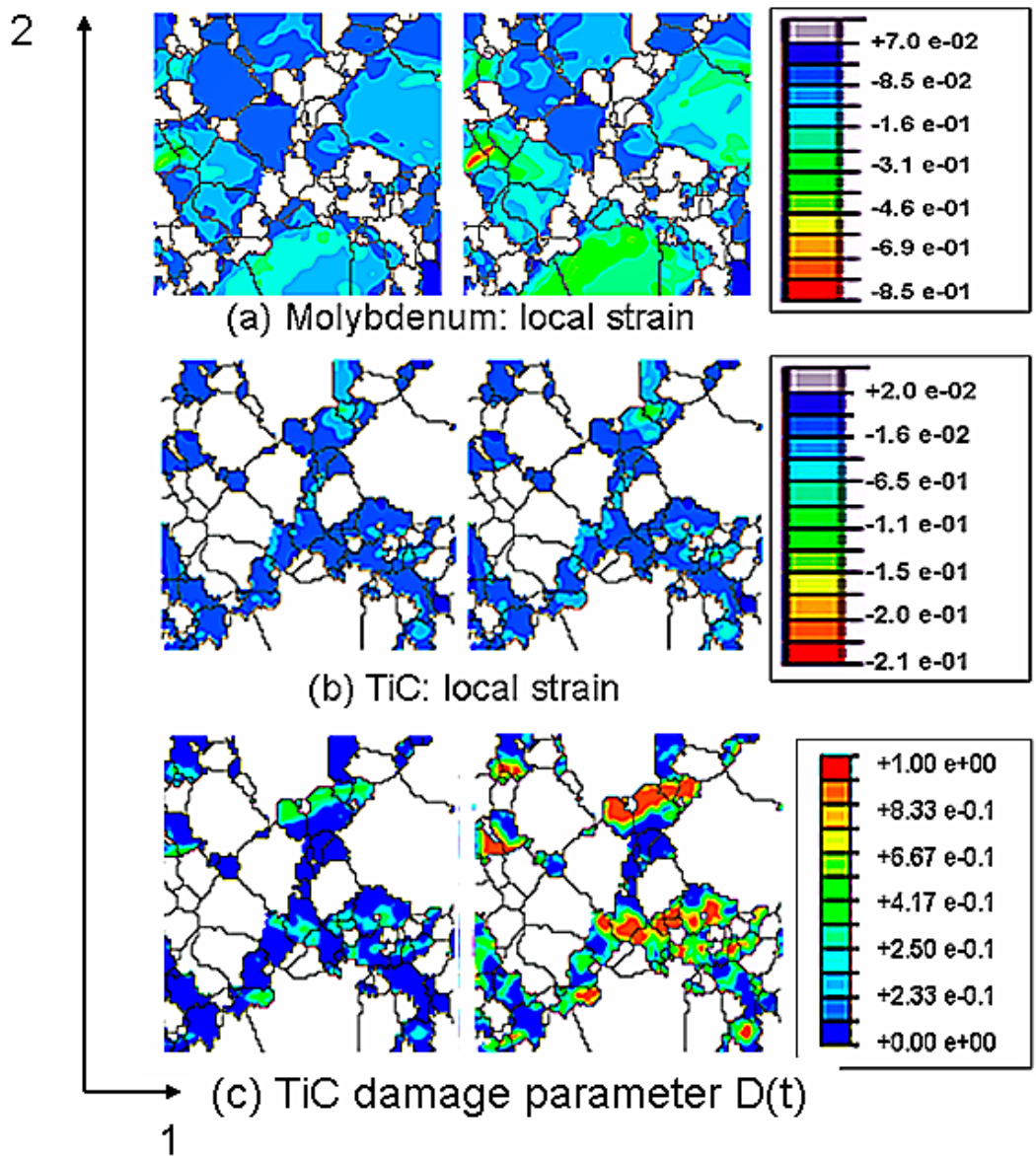

*Fig.6: Local strain* $\varepsilon_{22}$ *evolution under two applied strains E= $E_{22}$=-0.03 and E= $E_{22}$=-0.1 at room temperature ( $\bar{2}$ Axis corresponds to the compression axis). (a) Molybdenum, (b) TiC, (c) Damage parameter in TiC.*

5.2 Composite results for high temperature

The numerical simulation and the corresponding experimental curves for high temperatures are given in Fig.7.

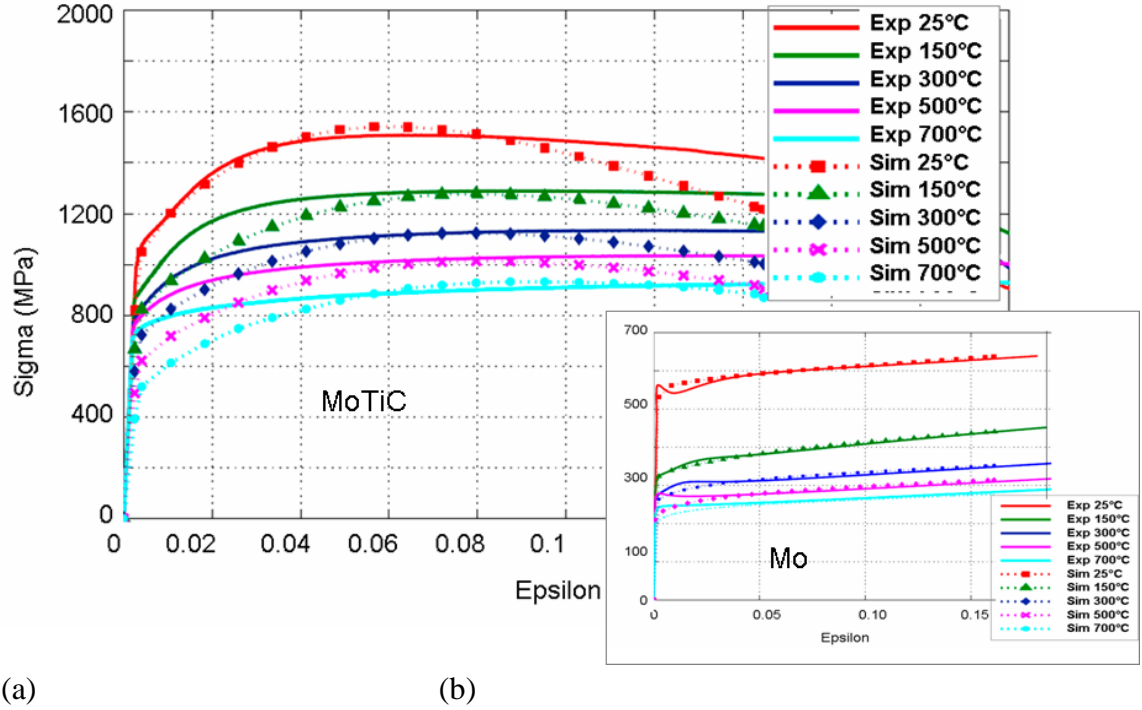

(a)                                        (b)

*Fig.7: Experimental and numerical stress – strain curves on a 3D numerical and a real polycrystalline aggregate of the composite for five temperatures. Experimental and numerical stress – strain curves on molybdenum are given in the insight.*

The comparison between the composite (Fig.7a) and the pure molydenum curves (Fig.7b) reveals, for an applied strain $E_{22} = -0.09$ , a large increase of the macroscopic yield stress form 600MPa to 1450MPa.

Fig. 7a shows, for $E_{22} = -0.09$ , a good agreement between numerical and experimental results. The difference $\Delta\Sigma_{22}$ between macroscopic experimental stress is less than *10MPa*. For small strains, the agreement is only qualitative.

For the composite, the small differences between experiment and simulation observed at the very beginning of plasticity stage come from the material parameters determined for pure molybdenum. For pure molydenum, the simulated yield stresses are lower than the experimental ones (Fig.7b). The material parameters of pure molydenum (tables 3, 4) were identified, for $E_{22} \geq 0.2\%$ , on simplified stress-strain curves. An accurate identification would need a higher numbers of material parameters. In addition, the damage may be considered as too much simple. In stage II, the average computed effective elastic modulus of TiC bridges decreases less than actually observed.

The computed distributions of internal stress $\sigma_{22}$ for TiC ($\sigma_{22}^{TiC}$) and for Mo ($\sigma_{22}^{Mo}$) are given Fig. 8a and Fig.8b. In order to understand the composite phase behaviour, the $\sigma_{22}^{Mo}$ distribution for a pure molydenum aggregate is reported in Fig.8c.

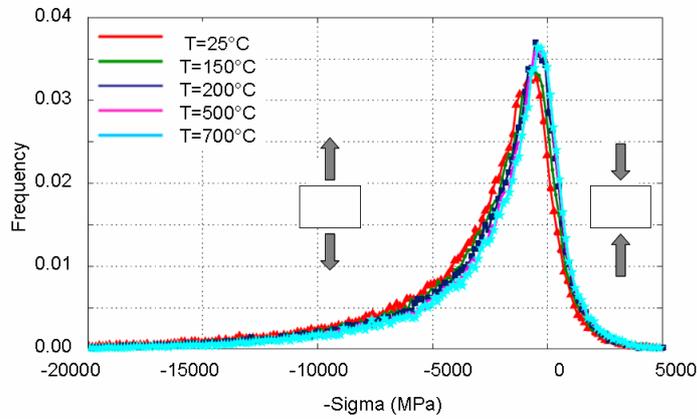

(a)

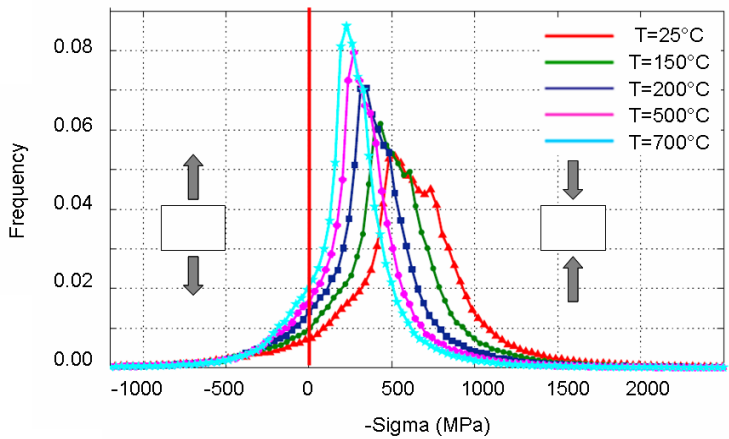

(b)

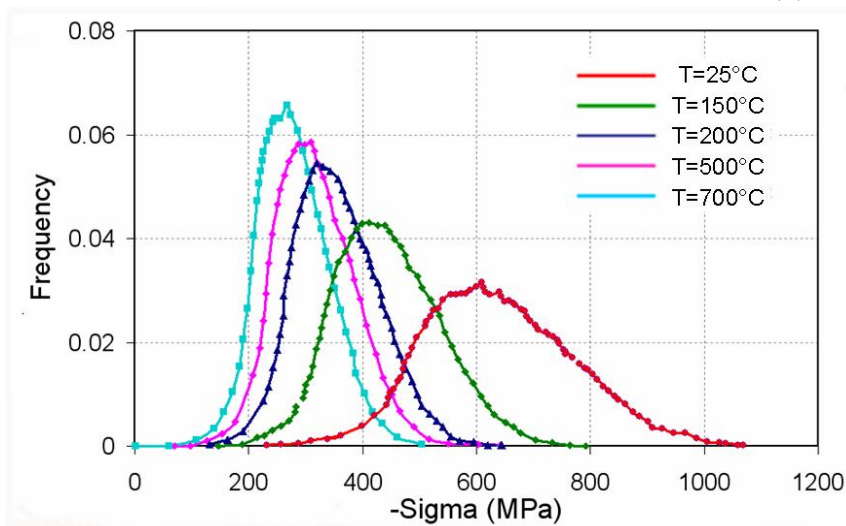

(c)

*Fig.8: Distribution of the uniaxial stress Sigma=$\sigma_{22}$ at E22=- 0.09. (a) TiC phase, (b) Molydenum matrix, (c) Pure molydenum*

Whatever the temperature, the TiC behaviour being assumed elastic and brittle, the distribution of $\sigma_{22}^{TiC}$ is independent of the matrix behaviour. On the contrary, in Mo matrix, $\sigma_{22\,C}^{Mo}$ distribution depends on both temperature and TiC behaviour.

*Molydenum matrix*: the comparison of $\sigma_{22\,C}^{Mo}$ (Mo matrix) and $\sigma_{22}^{Mo}$ (pure Mo) distributions for $E_{22} = -0.09$, raises several important points.

- The $\sigma_{22\,C}^{Mo}$ distributions present a large background noise and larger distributions than $\sigma_{22}^{Mo}$. It should be noted that some points of Mo matrix are actually in traction. In these points, the stress values can reach 500MPa and can lead to damage of the matrix.

- For low temperatures, the $\sigma_{22\,C}^{Mo}$ distribution curves presents an asymmetry shape which disappears with increasing temperature. Our computation reveals that for 25°C, 150°C and 300°C, molydenum grains present, close to inter-phase boundaries, very high internal stresses ($\sigma_{22\,C}^{Mo} \leq -500\,MPa$).

- The average values $\left\langle \sigma_{22\,C}^{Mo} \right\rangle$ and of $\left\langle \sigma_{22}^{Mo} \right\rangle$ are very close and present the same evolution with temperature. These values correspond to the macroscopic stress of pure molydenum compression curves.

*TiC carbide*. Though, for $E_{22} = -0.09$, some TiC bridges are broken (D(t)=1), a small part ($\approx 15\,\%$) of the TiC phase is still submitted to local tensile stresses greater than the critical stress ($\sigma_c \geq 250\,MPa$). Damage evolution is then slowed down in stage III: the skeleton of TiC presents a lot of micro-cracks and the composite is reduced to a Mo matrix with large TiC inclusions. As a result, the composite hardening slopes become constant (plateau) before presenting a negative slope. For high temperatures, the evolution with applied deformation of damage parameter, is slowed down compared to low temperature, leading to an increase of the size of the plateau

Damage is represented by a decay of the effective elastic modulus in some elements of the TiC mesh as soon as the criterion of the critical stress is satisfied. The square elements of the meshing are more representative of small holes, than actual thin micro-cracks. The increasing number of "holes" in TiC is sufficient to explain the weak negative slope of simulated stress-strain curves for large strain.

## 6. Conclusions

Understanding the mechanical behaviour of two-phase Mo-TiC30vol.% metal-ceramic composite, in a wide temperature range, requires simultaneous experimental and numerical studies. The first difficulty is to work on a realistic representation of the material, then to take into account of the damage of the brittle phase and its effect on the ductile one. The studied composite presents small grains and a brittle phase presenting percolation which crystalline orientation is not easily observable. Mechanical polishing and EBSD measurement cannot be used in this case and aggregate built by random layers of grains lead to wrong results. Consequently, a dual beam FIB-SEM was used to characterize and reconstruct layer by layer a real 3D polycrystalline two-phase Mo-TiC30vol.% metal-ceramic composite aggregate. Implementation of damage of the brittle phase at the scale of the grains is required and provides good results on the microscopic stress-strain fields when applied to this 3D aggregate. The TiC damage criterion is described in terms of an accumulative decay in the elastic stiffness of this phase as a function of straining. A very good agreement with experimental curves is observed at 25°C. For any temperature, the stage III behavior is accurately described.

The proposed 3D aggregate and the corresponding two-phase polycrystal model can be used to forecast the behavior of any related metal-ceramic or metal-metal composites at different temperatures, just by a change of the matrix and particles parameters.

**Acknowledgements**

The microstructure of the starting composite was characterized at the Max Planck Institute Düsseldorf. The authors are very grateful to Prof. S. Zaefferer for his kind and helpful discussions.

The studied composite MoTiC was elaborated at Comissariat à l'Energie Atomique. The authors thank Dr. M. Le Flem and Dr. J.L. Bechade to give us the material and for financial help for tomography experiments.

The authors are grateful to Prof. J.H. Schmitt (ECP) for stimulating and helpful discussions.

**Figure captions**

*Fig.1. SEM picture showing the microstructure of the Mo-TiC$_{30\%vol}$ composite revealing three types of phases, namely, Molybdenum (in light grey), TiC (in black), and (Mo,Ti)C (in grey) [1].*

*Fig.2: 3D representation of the phases obtained by X-ray tomography. (a) TiC , (b) Mo, (c) composite. Specimen size: (30 X 30 X 30 $\mu m^3$).*

*Fig.3: 30 X 30 X 6.6 $\mu m^3$ aggregate obtained by 3D electron orientation microscopy (3D EBSD). TiC and* TiC-Mo$_{15at.\%}$ *phases (top row), whole aggregate (bottom row). The colour code is given by the crystallographic standard triangle. For simulations, the loading direction is parallel to $\bar{2}$ axis.*

*Fig.4: Composite behavior compression experimental tests; sigma: tensile stress; F: force; S(t): cross section at each time interval ; $\varepsilon_{22}$ : strain.*

*Fig.5: Mo, TiC and Mo-TiC$_{30\%}$ numerical and experimental compression curves (T=25°C); sigma22: compression stress; epsilon22: strain.*

*Fig.6. Local strain $\varepsilon_{22}$ and damage evolution under two applied strains E= E$_{22}$=-0.03 and E= E$_{22}$=-0.1 at room temperature ( $\bar{2}$ Axis corresponds to the compression axis). (a) Molybdenum phase, (b) TiC phase, (c) damage parameter.*

*Fig.7. Experimental and numerical stress – strain curves on a 3D numerical and a real polycrystalline aggregate of the composite for five temperatures. Experimental and numerical stress – strain curves on molybdenum are given in the insight.*

*Fig.8. Distribution of the uniaxial stress Sigma= $\sigma_{22}$ for E22=- 0.09. (a)TiC phase, (b )Molydenum matrix, (c) Pure Molydenum.*

Molybdenum

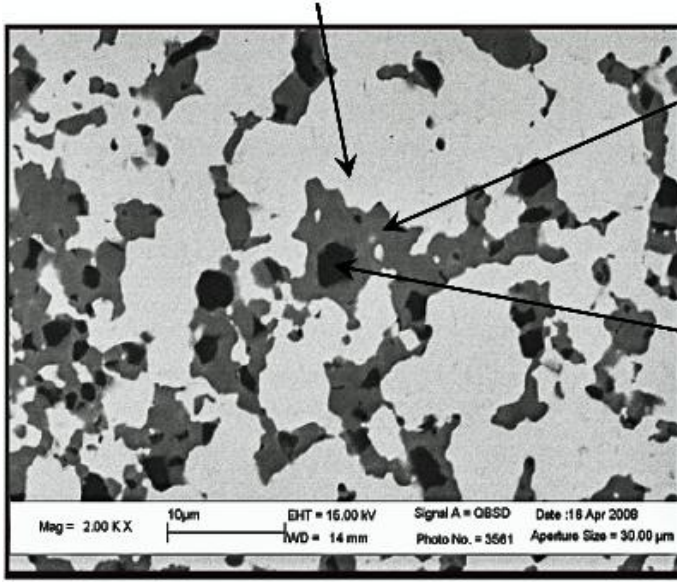

(Mo,Ti)C Phase

Titanium Carbide

*FIG1*

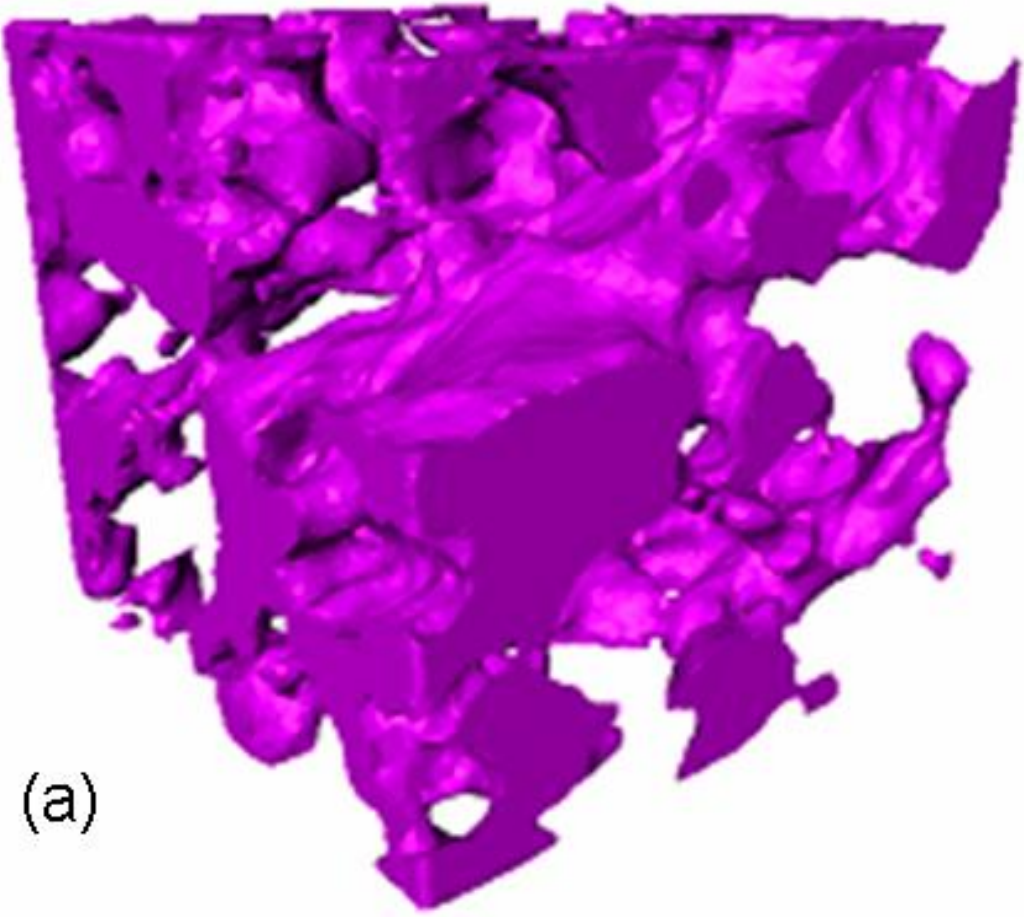

(a)

*FIG2a*

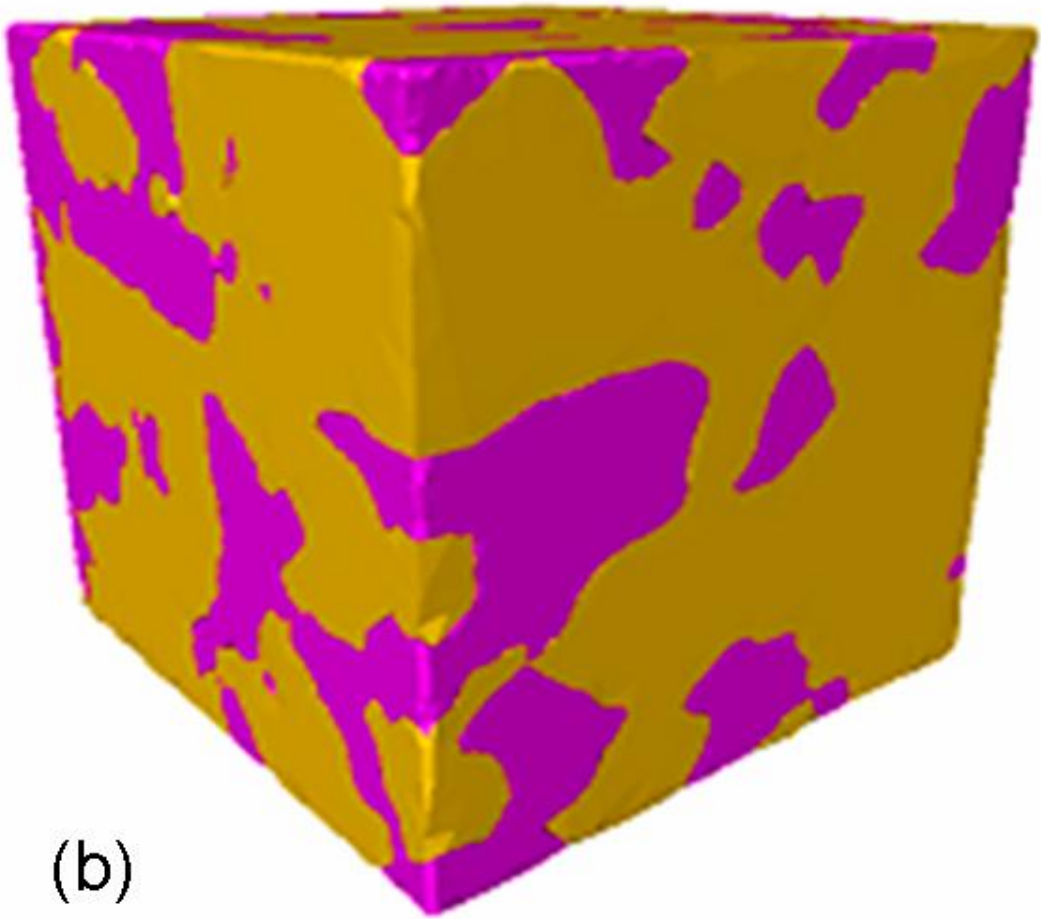

(b)

*FIG2b*

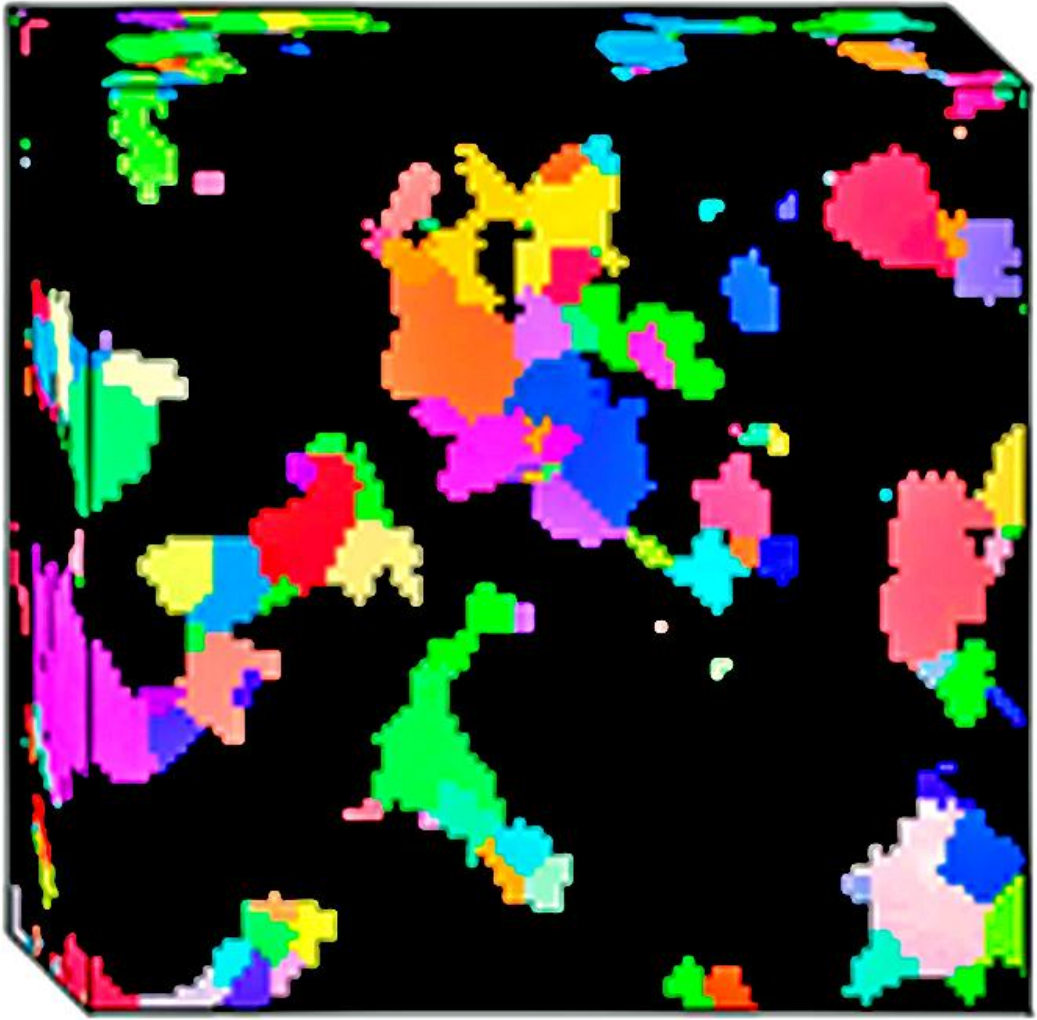

*FIG3a*

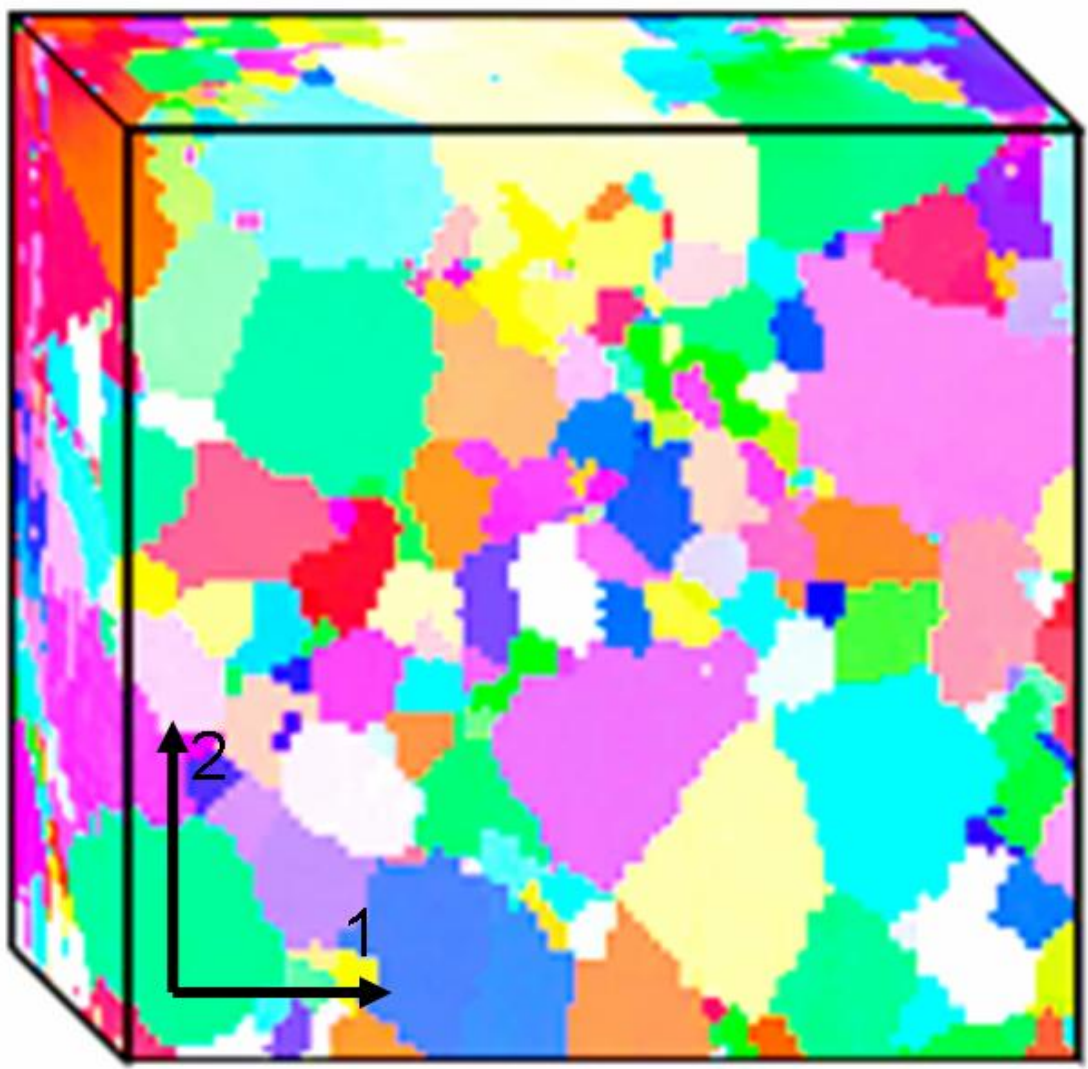

*FIG3b*

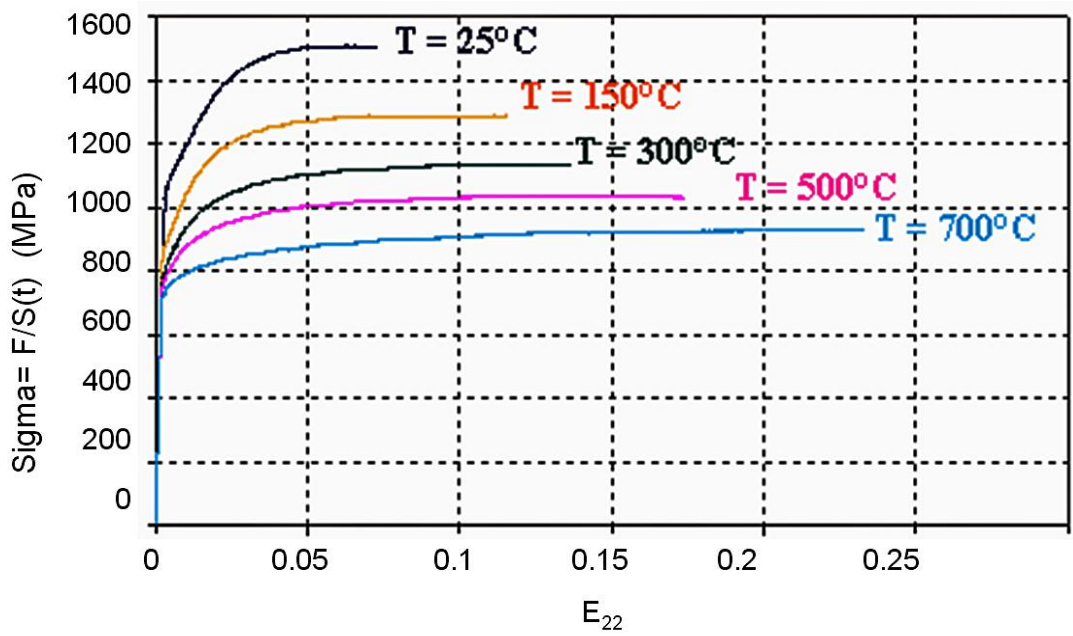

*FIG4*

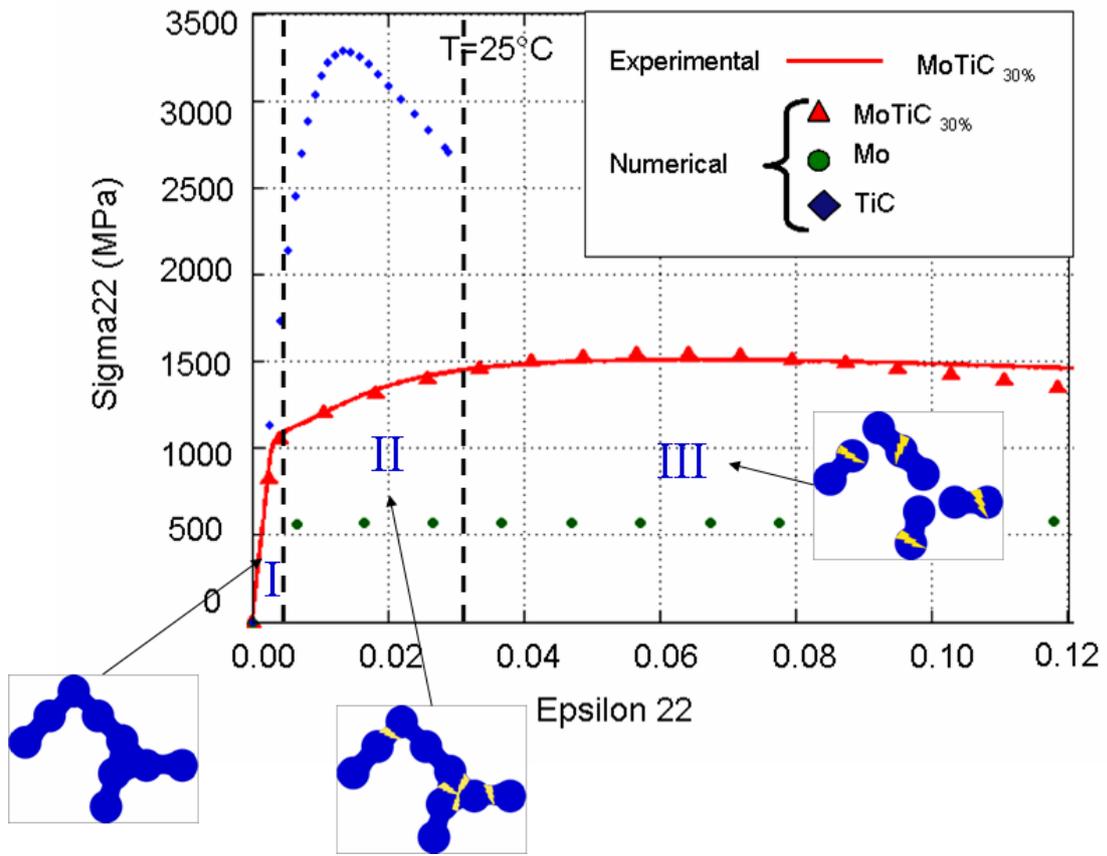

*FIG5*

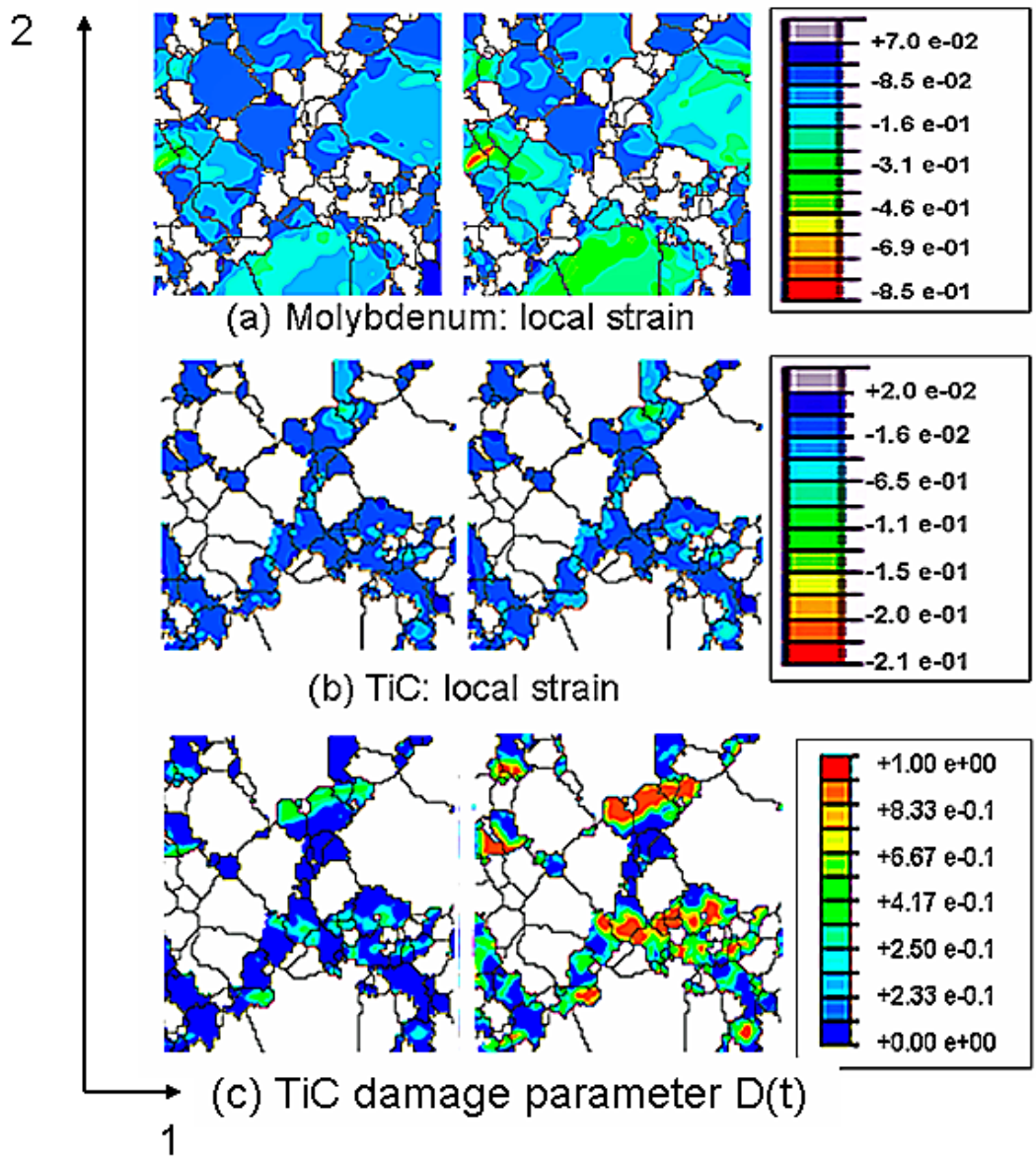

(a) Molybdenum: local strain

| +7.0 e-02 |
| -8.5 e-02 |
| -1.6 e-01 |
| -3.1 e-01 |
| -4.6 e-01 |
| -6.9 e-01 |
| -8.5 e-01 |

(b) TiC: local strain

| +2.0 e-02 |
| -1.6 e-02 |
| -6.5 e-01 |
| -1.1 e-01 |
| -1.5 e-01 |
| -2.0 e-01 |
| -2.1 e-01 |

(c) TiC damage parameter D(t)

| +1.00 e+00 |
| +8.33 e-0.1 |
| +6.67 e-0.1 |
| +4.17 e-0.1 |
| +2.50 e-0.1 |
| +2.33 e-0.1 |
| +0.00 e+00 |

*FIG6*

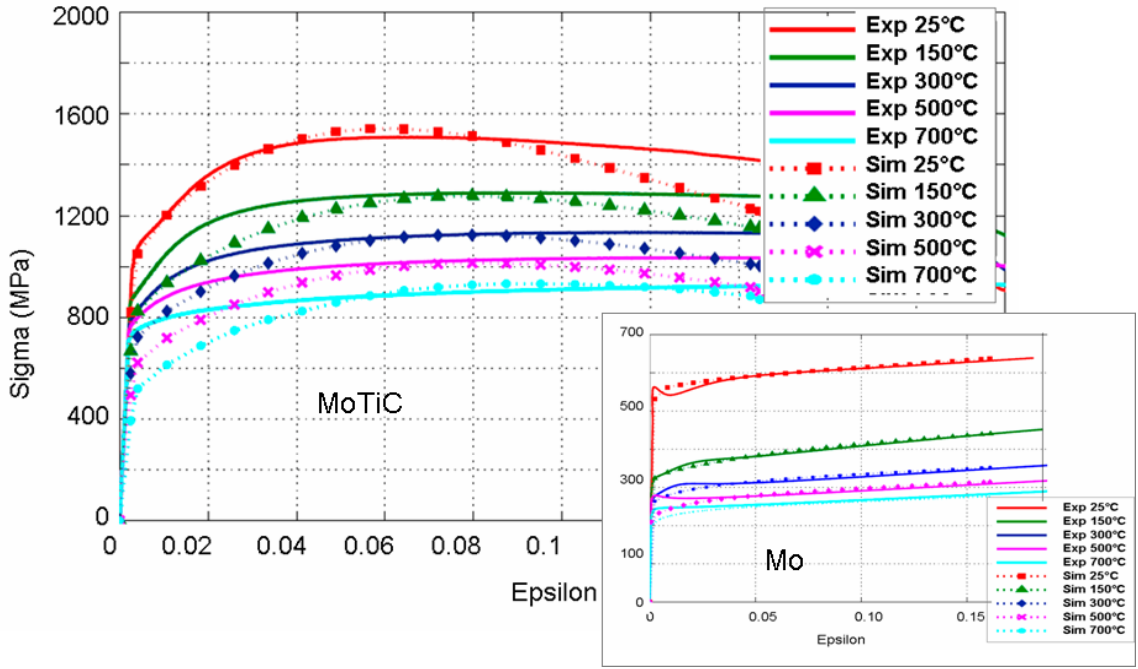

*FIG_7*

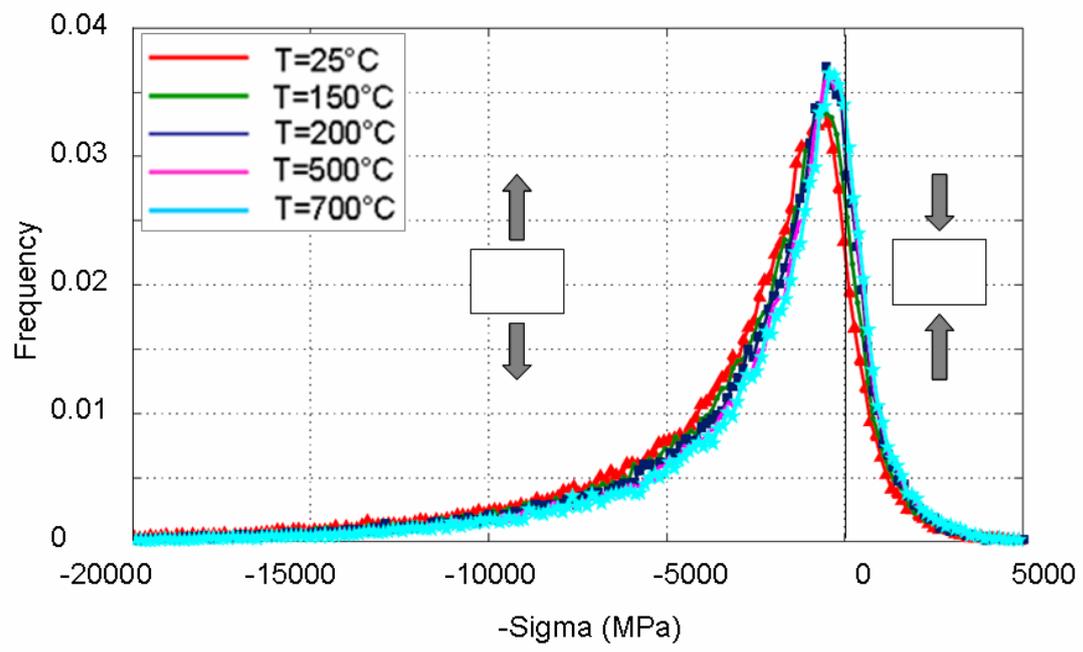

*FIG_8a*

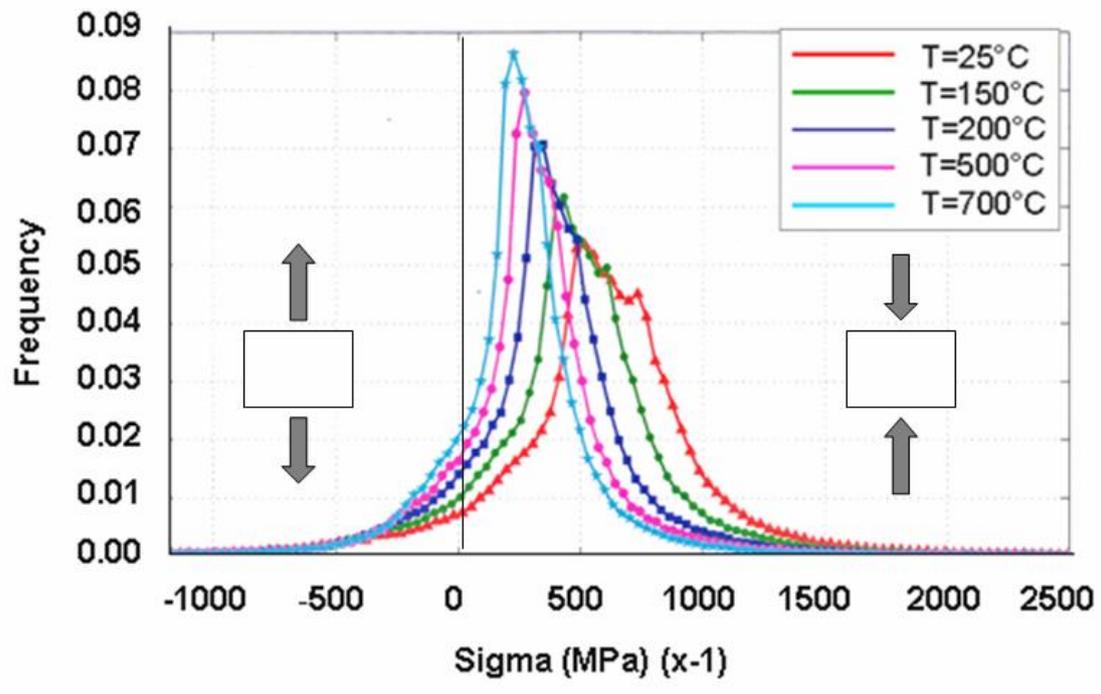

*FIG_8b*

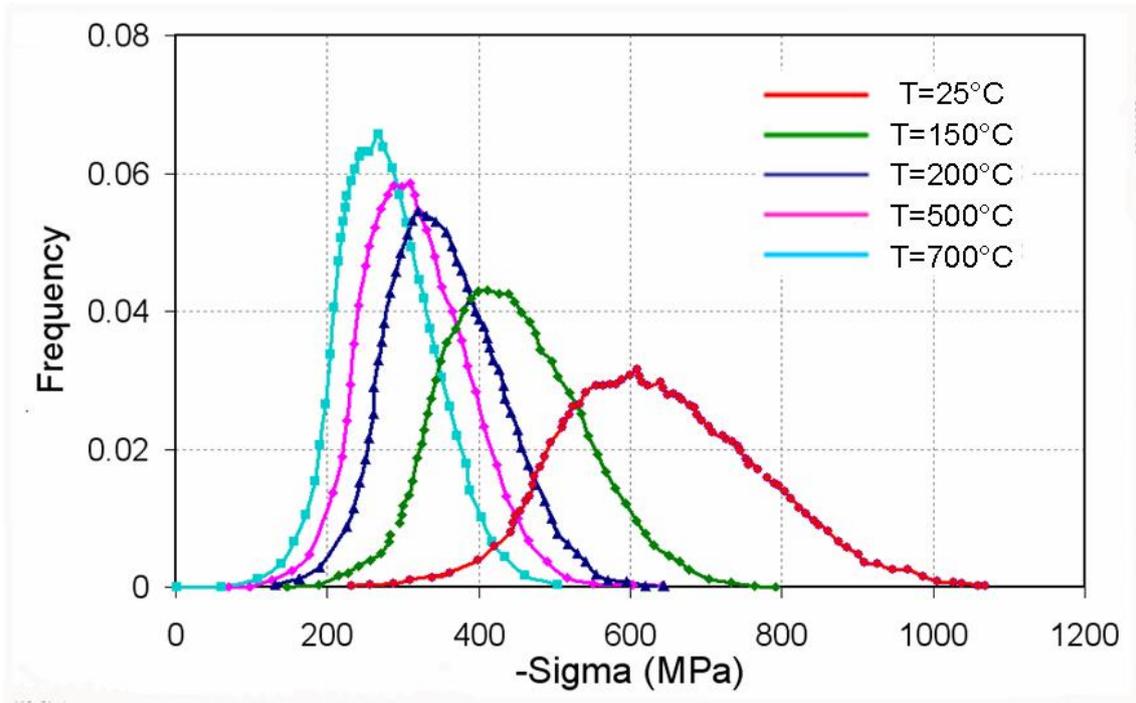

*FIG_8c*

Table 1. TiC composition (%mass).

| C | O | N | Ca | Co | W | Ni | Al | Fe | S |
|---|---|---|---|---|---|---|---|---|---|
| 19.23 | 0.6126 | 0.0279 | 0.002 | 0.032 | 0.39 | <4E-4 | 0.0014 | 0.0061 | 0.0019 |

Table 2. Molybdenum composition (% mass)

| Mo | O | Fe | K |
|---|---|---|---|
| 99.98 | 0.0620 | 9ppm | 29ppm |

Table 3. *Set of parameters identified for molybdenum by Cédat et al [1]*

| $\tau_0$ | $a^{uu}=a^{su}$ | $\rho_0$ | $g_{c0}$ | $E_{gc}$ | $D_{grain}$ |
|---|---|---|---|---|---|
| 125 MPa | 0.01 | $10^{-11}\mathrm{m}^{-2}$ | 14nm | $2.17\mathrm{x}10^{-2}$ eV | 3 µm |
| $\tau_R$ | $\dot{\gamma}_0$ | $\Delta G_0$ | p | q | |
| 498 MPa | $10^{-1}$ | 1.1 eV | 0.2 | 1.5 | |

Table 4. *Set of temperature dependant parameters identified for molybdenum*

| Molybdenum parameters depending on temperature | | | |
|---|---|---|---|
| T | $\Delta G(T)$ (eV) | $\tau_0$ (MPa) | K | $g_c$ (nm) |
| 25°C | 0.35 | 125 | 440 | 20 |
| 150°C | 0.5 | 70 | 80 | 12 |
| 300°C | 0.97 | 70 | 50 | 26 |
| 500°C | 1.18 | 70 | 50 | 30 |
| 700°C | - | 70 | 30 | 36 |

Table 5: Parameters of titanium carbide (TiC) for the proposed model

| E (GPa) | $\nu$ | $\alpha$ | $\beta$ | $\sigma_c$ (MPa) |
|---------|-------|----------|---------|------------------|
| 440 | 0.19 | 0.013 | 0.013 | 250 |

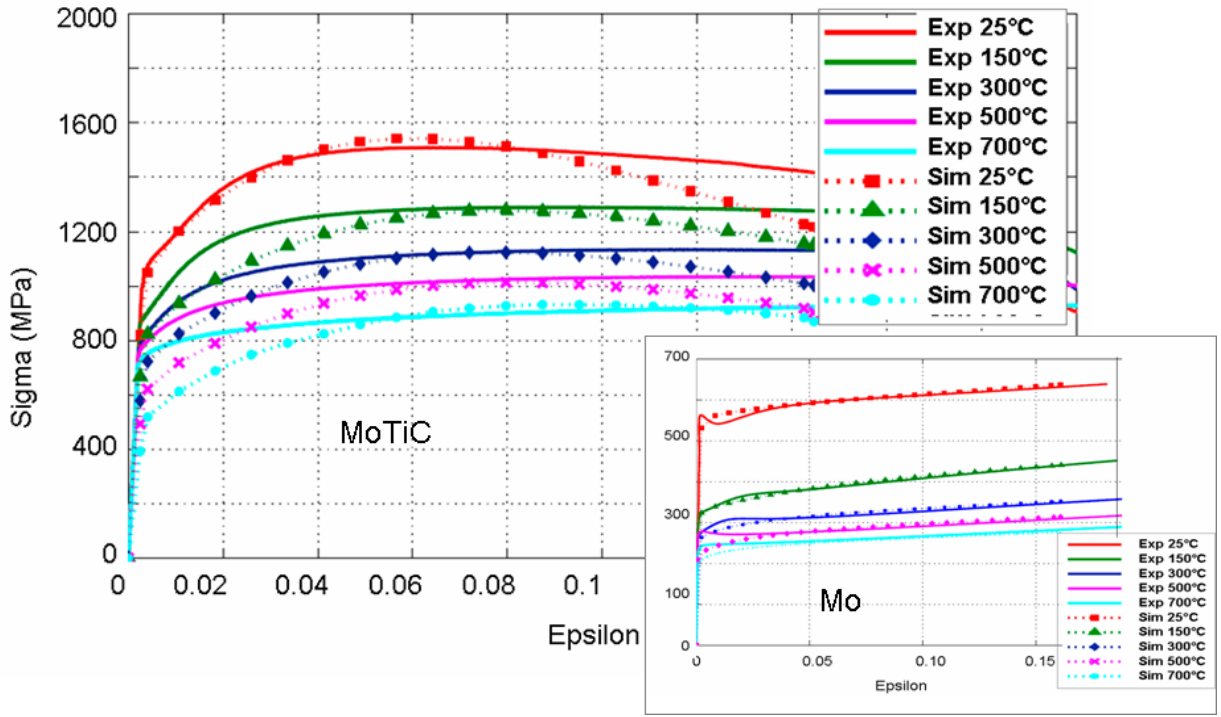

*FIG7*

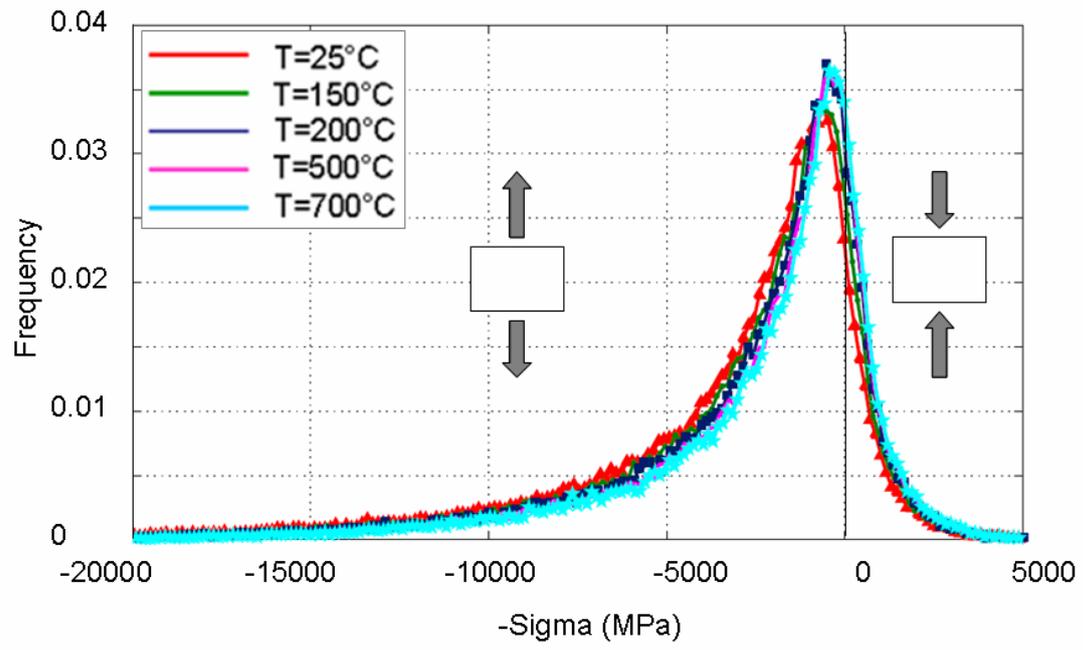

*FIG_8a*

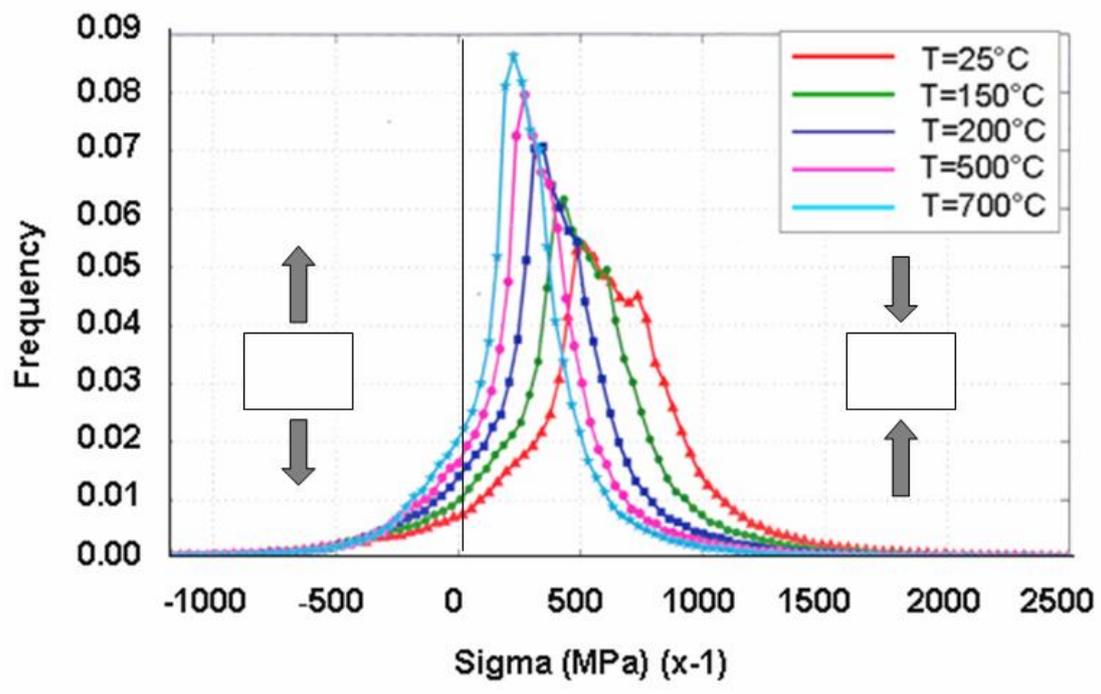

*FIG_8b*

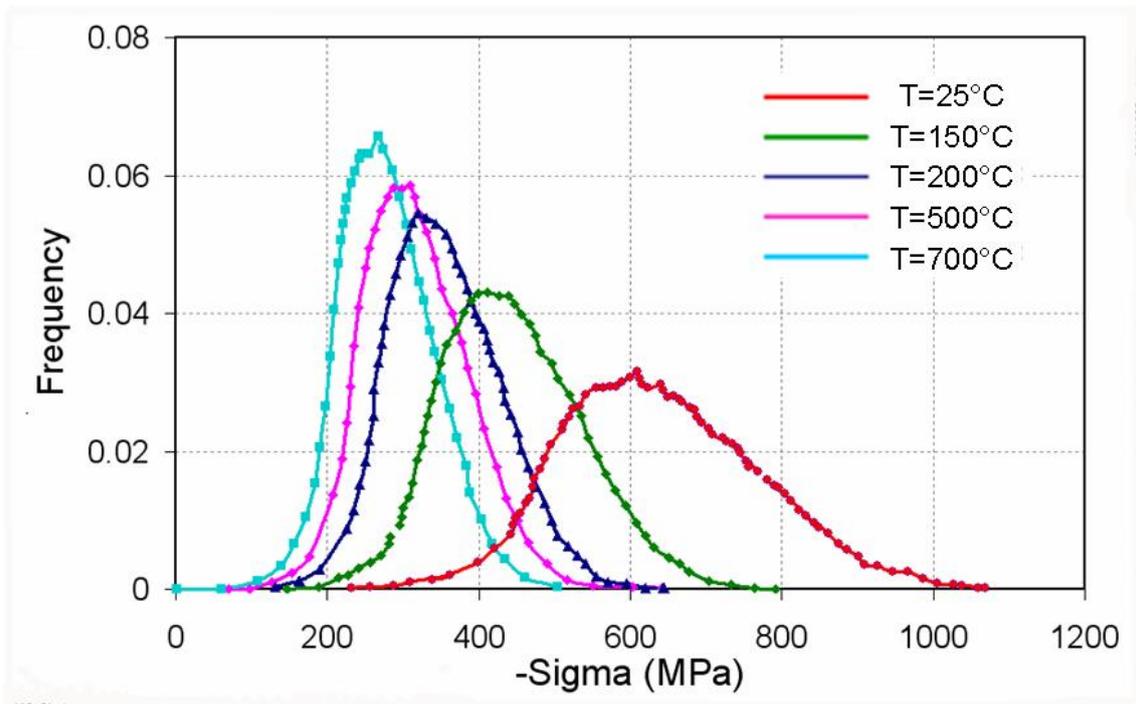

*FIG_8c*

Table 1. TiC composition (%mass).

| C | O | N | Ca | Co | W | Ni | Al | Fe | S |
|---|---|---|---|---|---|---|---|---|---|
| 19.23 | 0.6126 | 0.0279 | 0.002 | 0.032 | 0.39 | <4E-4 | 0.0014 | 0.0061 | 0.0019 |

Table 2. Molybdenum composition (% mass)

| Mo | O | Fe | K |
|---|---|---|---|
| 99.98 | 0.0620 | 9ppm | 29ppm |

Table 3. *Set of parameters identified for molybdenum by Cédat et al [1]*

| $\tau_0$ | $a^{uu}=a^{su}$ | $\rho_0$ | $g_{c0}$ | $E_{gc}$ | $D_{grain}$ |
|---|---|---|---|---|---|
| 125 MPa | 0.01 | $10^{-11}\mathrm{m}^{-2}$ | 14nm | $2.17\mathrm{x}10^{-2}$ eV | 3 μm |
| $\tau_R$ | $\dot{\gamma}_0$ | $\Delta G_0$ | p | q | |
| 498 MPa | $10^{-1}$ | 1.1 eV | 0.2 | 1.5 | |

Table 4. *Set of temperature dependant parameters identified for molybdenum*

| \multicolumn{5}{l}{Molybdenum parameters depending on temperature} |
| T | $\Delta G(T)$ (eV) | $\tau_0$ (MPa) | K | $g_c$ (nm) |
|---|---|---|---|---|
| 25°C | 0.35 | 125 | 440 | 20 |
| 150°C | 0.5 | 70 | 80 | 12 |
| 300°C | 0.97 | 70 | 50 | 26 |
| 500°C | 1.18 | 70 | 50 | 30 |
| 700°C | - | 70 | 30 | 36 |

Table 5: Parameters of titanium carbide (TiC) for the proposed model

| E (GPa) | $\nu$ | $\alpha$ | $\beta$ | $\sigma_c$ (MPa) |
|---------|-------|----------|---------|------------------|
| 440 | 0.19 | 0.013 | 0.013 | 250 |